\documentclass[sigconf]{acmart}
\AtBeginDocument{%
  \providecommand\BibTeX{{%
    \normalfont B\kern-0.5em{\scshape i\kern-0.25em b}\kern-0.8em\TeX}}}

\copyrightyear{2022}
\acmYear{2022}
\setcopyright{acmlicensed}
\acmConference[ICAIF '22]{3rd ACM International Conference on AI in Finance}{November 2--4, 2022}{New York, NY, USA}
\acmBooktitle{3rd ACM International Conference on AI in Finance (ICAIF '22), November 2--4, 2022, New York, NY, USA}
\acmPrice{15.00}
\acmDOI{10.1145/3533271.3561755}
\acmISBN{978-1-4503-9376-8/22/10}

\usepackage{caption}
\usepackage{subcaption}
\usepackage{graphicx}
\usepackage{amsmath}
\usepackage{amsthm}
\usepackage{booktabs}
\usepackage{algorithm}
\usepackage{algorithmic}
\usepackage{xcolor}

\newcommand{\R}{\mathbb{R}}
\newcommand{\tX}{\tilde{X}}
\newcommand{\tY}{\tilde{Y}}
\newcommand{\tn}{\tilde{n}}
\newcommand{\cX}{\mathcal{X}}
\newcommand{\bmu}{\bar{\mu}}
\newcommand{\bsigma}{\bar{\sigma}}

\begin{document}

\title{Efficient Calibration of Multi-Agent Simulation Models from Output Series with Bayesian Optimization}


\author{Yuanlu Bai}
\email{yb2436@columbia.edu}
\author{Henry Lam}
\email{henry.lam@columbia.edu}
\affiliation{%
  \institution{Columbia University}
  \city{New York}
  \state{New York}
  \country{USA}
}
\author{Tucker Balch}
\email{tucker.balch@jpmchase.com}
\author{Svitlana Vyetrenko}
\email{svitlana.s.vyetrenko@jpmchase.com}
\affiliation{%
  \institution{J.P.Morgan AI Research}
  \city{New York}
  \state{New York}
  \country{USA}
}


\begin{abstract}
Multi-agent simulation is commonly used across multiple disciplines, specifically in artificial intelligence in recent years, which creates an environment for downstream machine learning or reinforcement learning tasks. In many practical scenarios, however, only the output series that result from the interactions of simulation agents are observable. Therefore, simulators need to be calibrated so that the simulated output series resemble historical -- which amounts to solving a complex simulation optimization problem. In this paper, we propose a simple and efficient framework for calibrating simulator parameters from historical output series observations. First, we consider a novel concept of eligibility set to bypass the potential non-identifiability issue. Second, we generalize the two-sample Kolmogorov-Smirnov (K-S) test with Bonferroni correction to test the similarity between two high-dimensional distributions, which gives a simple yet effective distance metric between the output series sample sets. Third, we suggest using Bayesian optimization (BO) and trust-region BO (TuRBO) to minimize the aforementioned distance metric. Finally, we demonstrate the efficiency of our framework using numerical experiments both on a multi-agent financial market simulator.
\end{abstract}



\maketitle

\section{Introduction} \label{sec:introduction}

Multi-agent simulation is commonly used across multiple disciplines to model counterfactual scenarios and support  decision-making. With rapid development of artificial intelligence, simulation is also applied to this area in recent years, which sets up an infrastructural environment where machine learning or reinforcement learning experiments can be carried out. For instance, experimentation with trading strategies in ``live'' environments can be costly or otherwise impossible, and therefore, before they are released to production, trading strategies and algorithms need to undergo extensive testing inside simulated environments under a variety of markets scenarios. Additionally, it is also the case that realistic simulated environments are needed for training trading strategies using reinforcement learning. To ensure that the simulation tool is useful, it is of vital importance to calibrate simulator parameters accurately, as the simulation model can be misleading if it is far off from reality. A peculiar calibration challenge is that in practical scenarios, it is often the case that the agent-level data is unavailable, and only the output series are directly observable. As an example, in multi-agent financial market simulation, usually only the output time series that result from interaction of multiple market agents (such as price and volume time series) are available to the general public. 

In this paper, our main contribution is to propose an efficient learning-based framework to calibrate the simulator parameters only from such output-level series samples. We note that the methodology can be applied to calibrate large-scale or even black-box simulation models in general, as the methodology does not assume specific structure of the simulators. The framework can be applied as long as the simulator generates output series from a probability distribution determined by the input parameters and real output data are observable.




Large-scale simulation systems usually have many parameters that are necessary to capture the complexity. Calibrating a simulator with respect to a large number of parameters can potentially cause the non-identifiability issue~\cite{tarantola2005inverse} -- namely, a different configuration may have the same or very similar output distribution as the true configuration, so they are indistinguishable only from the output data. To address this issue, we employ a novel concept of eligibility set~\cite{bai2020calibrating,bai2021calibrating}, where we measure the discrepancy between the output sample distributions instead of the input parameter values and then accept or reject the candidate configuration using the idea of hypothesis testing. To be specific, with a distance metric between output series sample sets as well as the critical value chosen accordingly, the candidate is not rejected if and only if the distance metric is below the critical value.


To define the distance metric and the critical value properly, we generalize the two-sample Kolmogorov-Smirnov (K-S) test~\cite{hodges1958significance} with Bonferroni correction \cite{shaffer1995multiple} to test whether two series sample sets are from the same distribution. As a distance metric, the K-S statistic is easy and convenient to compute, and the corresponding critical value with statistical guarantees can be computed or approximated explicitly. Before computing the K-S statistic, feature extraction can also be applied on the series samples in order to reduce the sample dimension and hence reduce the critical value conservativeness. For simplicity, we do not apply feature extraction in our experiments, which turns out to perform well under our experimental setups. We will discuss feature extractors using stylized facts~\cite{vyetrenko2019get,cont_lob} and learning-based methods~\cite{bai2021calibrating,masgan} and analyze their pros and cons in Appendix~\ref{app:feature_extractor}.


In this paper, we choose the K-S statistic between the real and simulated series (i.e. series that result from simulation) as an objective function for simulator calibration. Multi-agent simulator calibration poses a difficult optimization problem. First, with hundreds of or thousands of (or even more) agents interacting in the system, the simulation tends to be slow, so the objective function is expensive and thus the budget of evaluations can be extremely limited. Second, the objective function is likely to lack both special structure (such as convexity) and derivative information, which impedes us from leveraging them to improve efficiency. Third, the objective function is noisy as the K-S statistic depends on random samples. To combat the above issues, we suggest using Bayesian optimization(BO)~\cite{jones1998efficient,frazier2018tutorial}, which has been a popular and competitive approach to optimize expensive black-box objective functions in recent years. However, it is well-known that BO does not scale well to high-dimensional cases, and therefore, we recommend a state-of-the-art variant called trust region BO (TuRBO)~\cite{eriksson2019scalable}, which adopts a local strategy instead.


To the best of our knowledge, we are the first to use the generalized K-S statistic as a distance between high-dimensional series sample sets to calibrate multi-agent simulators. We are aware of its limitation of not using series properties such as serial dependence information. We, however, believe that this approach is justified for the purpose of calibrating simulators instead of modeling the series. First, it enjoys statistical guarantees on correctness and conservativeness~\cite{bai2021calibrating}, which supports its effectiveness in high-dimensional usage. Second, compared to learning-based metrics, this statistic is more explainable, and also easier to evaluate without needing to be trained. Third, after eliminating the random noise, it is a relatively smooth function of the input configuration and thus enables us to use the Gaussian process (GP) surrogate model in BO algorithm, which is not only simple but also able to provide interval estimation besides point prediction. In the literature, BO has been applied to calibrate agent-based models, but the objective function and the surrogate model are often chosen differently. \cite{tran2020bayesian} is the closest work to ours that we have found. However, they use the one-dimensional K-S statistic between daily return distributions, whereas we use a high-dimensional generalization of K-S statistic to compare the entire series distributions. They also choose the tree-structured Parzen estimator as the surrogate model instead of GP. 

From our numerical results, we observe that despite its simplicity, combining the K-S statistic distance metric and the GP surrogate works well in practice. We foresee our calibration framework would be of interest and adaptable to other problems in the broad field of multi-agent simulation. 


\section{Calibration Framework} \label{sec:framework}

\subsection{Eligibility Set} \label{sec:eligibility_set}

Consider a simulation model which takes an input configuration $\theta\in\Theta\subset\R^I$ and outputs a series sample in $\R^O$ from distribution $P^{\theta}$. Suppose that the real distribution is $P^{real}$, then ideally our goal is to find $\theta\in\Theta$ such that $P^{\theta}=P^{real}$, or equivalently, $d(P^{\theta},P^{real})=0$ for any valid statistical distance $d(\cdot,\cdot)$ between probability distributions. However, the simulation model is not necessarily identifiable. That is, it is possible that there exist different configurations $\theta_1\ne\theta_2$ such that $$d(P^{\theta_1},P^{real})=d(P^{\theta_2},P^{real})=0.$$ 
In this case, only given the output-level information, it is impossible to identify which one is indeed the ``truth''. Therefore, instead of trying to recover the ``best'' configuration, we jointly consider the set $$\{\theta\in\Theta:d(P^{\theta},P^{real})=0\}.$$

However, in practice, $P^{real}$ is unknown and $P^{\theta}$ is hard to derive due to the model complexity. Thus, we can only estimate them using finite samples. Suppose real samples $$X_1,\dots,X_N\sim P^{real}$$ are available, and for each $\theta$, we generate samples $$Y_1^{\theta},\dots,Y_n^{\theta}\sim P^{\theta}$$ using the simulator. Then we can use the empirical distributions $$P_N^{real}(\cdot)=\frac1N\sum_{i=1}^N\delta_{X_i}(\cdot)$$
and 
$$P_n^{\theta}(\cdot)=\frac1n\sum_{j=1}^n\delta_{Y_j^{\theta}}(\cdot)$$ 
to approximate $P^{real}$ and $P^{\theta}$ respectively. As a relaxation, we construct the statistically confident eligibility set as 
$$\{\theta\in\Theta:d(P_n^{\theta},P_N^{real})<q\}$$ 
where $q\in\R^+$ is a suitable constant threshold. In particular, when $q$ is chosen as the critical value with significance level $\alpha$, the eligibility set is actually a $(1-\alpha)$-level confidence region for the ``true configuration''. We refer to~\cite{bai2021calibrating} for more details about the construction and theoretical guarantees of eligibility set.

\subsection{Choice of Distance Metric} \label{sec:distance_metric}

The choice of distance metric $d$ is critical for the success of calibration framework. First of all, given two sample sets from $P^{real}$ and $P^{\theta}$ respectively, the distance $d(P_n^{\theta},P_N^{real})$ should be easy to compute. Moreover, the choice of $d$ affects the conservativeness which is intuitively measured by the ``size'' of the eligibility set. As an extreme example, if $d(P_n^{\theta},P_N^{real})\equiv 0$, then the eligibility set is extremely conservative as it contains any $\theta\in\Theta$. Thus, we aim to pick an easily-evaluated distance metric which well captures the discrepancy between two series sample sets. 

In general, we consider first applying a feature extractor $f:\R^O\to\R^K$ on each series sample. That is, we compare $K$-dimensional sample sets $$\tX_i=f(X_i),i=1,\dots,N$$ and $$\tY_j^{\theta}=f(Y_j^{\theta}),j=1,\dots,n.$$ 
Note that setting $f$ as the identity map $x\mapsto x$ implies no feature extraction. A good choice of feature extractor should reduce the dimension of the original series without losing too much information.

Recall that the one-dimensional two-sample K-S test is a non-parametric test for equality of two distributions. To be specific, for any $k=1,\dots,K$, we use $\tX_{i,k}$ and $\tY_{j,k}^{\theta}$ to denote the $k$-th component of $\tX_i$ and $\tY_j^{\theta}$, and then $\tX_{i,k}$'s, $\tY_{j,k}^{\theta}$'s are two one-dimensional sample sets. Denote their empirical CDF as $F_{N,k}$ and $F_{n,k}^{\theta}$ respectively, i.e. 
$$
\begin{aligned}
&F_{N,k}(x)=\frac1N\sum_{i=1}^NI(\tX_{i,k}\leq x),\\
&F_{n,k}^{\theta}(x)=\frac1n\sum_{j=1}^nI(\tY_{j,k}^{\theta}\leq x).
\end{aligned}
$$ 
In the one-dimensional K-S test, the statistic 
$$\sup_{x\in\R}|F_{N,k}(x)-F_{n,k}^{\theta}(x)|$$ 
is contrasted with the critical value $q_{N,n,\alpha}$. If the statistic is greater than the critical value, then we reject the null hypothesis that $\tX_{i,k}$'s and $\tY_{j,k}^{\theta}$'s come from the same distribution with significance level $\alpha$. 

The K-S test cannot be easily generalized to high-dimensional case since the joint CDF needs to be considered. As a substitute, we propose pooling the discrepancies in each dimension with Bonferroni correction. That is, the high-dimensional K-S statistic is defined as $$\max_{k=1,\dots,K}\sup_{x\in\R}|F_{N,k}(x)-F_{n,k}^{\theta}(x)|$$
and the $\alpha$-level critical value is defined as $q_{N,n,\alpha/K}$. Under the null hypothesis $P^{\theta}=P^{real}$, we have that $$
\begin{aligned}
&P(\max_{k=1,\dots,K}\sup_{x\in\R}|F_{N,k}(x)-F_{n,k}^{\theta}(x)|>q_{N,n,\alpha/K})\\ \leq&\sum_{k=1}^K P(\sup_{x\in\R}|F_{N,k}(x)-F_{n,k}^{\theta}(x)|>q_{N,n,\alpha/K})\\ \leq &\sum_{k=1}^K \frac{\alpha}{K}=\alpha.
\end{aligned}
$$
Hence, if the K-S statistic exceeds the critical value, then we can reject the null hypothesis with significance level $\alpha$.

Thus, we suggest choosing the K-S statistic as the distance metric $d(P_n^{\theta},P_N^{real})$ with $q=q_{N,n,\alpha/K}$. For simplicity, it is known that $q$ can be approximated with $$q\approx\sqrt{-\frac{(N+n)\log(\alpha/2K)}{2Nn}}.$$ 
From the above derivation, we know that a bad choice of feature extractor $f$ does not harm the correctness of the eligibility set. That is, the eligibility set defined in this way is always a $(1-\alpha)$-level ``confidence region'' for the true configuration (if any). However, as we have explained, the choice of $f$ affects the conservativeness. We will discuss different types of feature extractors in Appendix~\ref{app:feature_extractor} and we refer to~\cite{bai2021calibrating} for statistical guarantees on both correctness and conservativeness.

Finally, we explain why we do not choose the more widely known Kullback–Leibler (K-L) divergence as the distance metric. In fact, the difficulty of estimating the K-L divergence (or similar likelihood-based metrics) is a motivation for us to propose using the K-S statistic. More specifically, K-L divergence requires information on the likelihood ratio between the two probability distributions in consideration. For simulation model outputs, direct likelihood information is unavailable as the outputs are specified through system dynamics. Thus, to empirically estimate the K-L divergence, we either discretize the distributions or use variational representation. In both cases, another level of approximation error is incurred and extra tuning effort is needed. In \cite{sriperumbudur2010non}, it is pointed out that even for some well-designed consistent K-L estimators, the convergence rate can be arbitrarily slow, and thus the authors advocate integral probability metrics instead. The K-S distance that we propose belongs precisely to this latter class.

\subsection{Bayesian Optimization} \label{sec:bayesian_optimization}

For a continuous configuration space $\Theta\subset\R^I$, we cannot generate a sample set from each $\theta\in\Theta$ and check whether it is rejected or not. Thus, the eligibility set can only be approximated with discrete points. If we find some points in the eligibility set, then we can jointly consider them in downstream tasks. That is, to estimate a target quantity of $P^{real}$, instead of evaluating it only with the ``best'' configuration, we use all the ``eligible'' configurations to provide a more robust estimation (see~\cite{bai2021calibrating}), which enables us to bypass the non-identifiability issue. In the low-dimensional case such as $I=1,2$, the eligibility set can be approximated by grid search or random search. 

However, the parameter space grows exponentially in the number of parameters $I$. In order to find an eligible point in the higher-dimensional case, we consider minimizing the distance metric, i.e. the K-S statistic, over the entire parameter space. As mentioned in Section~\ref{sec:introduction}, this global optimization problem is hard, since the objective function is expensive, black-box and noisy. To solve such problems, BO has gained popularity in recent years. 

Before introducing BO, we first discuss some alternative commonly-used optimization algorithms. In our early-stage exploration, we had also tried to use various stochastic gradient descent algorithms, including finite difference \cite{bhatnagar2013stochastic}, variational optimization \cite{staines2012variational} and simultaneous perturbation stochastic approximation (SPSA) \cite{bhatnagar2013stochastic}, before we finally chose to use BO in our numerical experiments. It turned out that finite difference and variational optimization are not efficient enough under our settings since the evaluation of the objective function is so costly that we could not afford to evaluate the function values at many points to estimate the gradient in each iteration. While SPSA can be more computationally efficient than the other two approaches, it also did not perform well in our trial, and we think the possible reasons are (i) our objective function is relatively noisy and can have many local optima, and (ii) it can be sensitive to the selection of hyperparameters. Some recent works also show that BO performs better than SPSA in similar scenarios (e.g. \cite{sha2020applying}). Therefore, after a careful consideration, we chose to focus on using BO and improving its performance with the state-of-the-art variant TuRBO in this work.

The key idea of standard BO is to model the objective function as a realization of a prior GP model. After evaluating the objective values at some points, we gain some information on the objective function, and conditional on the information, the posterior GP model can be used to predict the objective function at any point. In the algorithm, we evaluate the objective values on randomly sampled initial points to gain initial information, and then repeatedly select the next point, evaluate the objective value and update the posterior model. 

Now we recap the idea of BO more technically. Consider the optimization problem $\min_{x\in\cX} g(x)$ where $g:\cX\to\R$ is the objective function. In our settings, $x$ is $\theta$, $\cX$ is $\Theta$ and $g$ is the distance metric $\theta\mapsto d(P_n^{\theta},P_N^{real})$. The key idea of standard BO is to model $g(x),x\in\cX$ as a realization of a GP. Note that this prior GP model is uniquely defined by its mean structure $$\mu(x):=E(g(x)),x\in\cX$$ and covariance structure $$\Sigma(x,x'):=cov(g(x),g(x')),x,x'\in\cX.$$ 

We fix the initial budget $n_{init}$ and the total budget $n_{total}$ in advance. The BO algorithm is initialized by randomly sampling $n_{init}$ points $x_1,\dots,x_{n_{init}}$ in $\cX$ and evaluating their objective values $g(x_1),\dots,g(x_{n_{init}})$. At each iteration, suppose we have already evaluated the objective values at $x_1,\dots,x_{\tn}$. In the prior model, for any $x$, we know that $(g(x_1),\dots,g(x_{\tn}),g(x))$ follows a Gaussian distribution. Now conditional on the values of $g(x_1),\dots,g(x_{\tn})$, we get that $$g(x)\sim N(\bmu(x),\bsigma^2(x))$$ where the posterior mean and variance function $\bmu$ and $\bsigma^2$ can be explicitly computed. In particular, the posterior model $$g(x)|(g(x_1),\dots,g(x_{\tilde{n}})),x\in\cX$$ is still a GP. If the total budget is reached, i.e. $\tn=n_{total}$, then we stop the algorithm and use the posterior GP model as a predictor for the objective function. Otherwise, with the posterior model, the next point $x_{\tn+1}$ is selected by optimizing the acquisition function, which can be interpreted as maximizing a reward function, and then $g(x_{\tn+1})$ is computed. In summary, we iteratively select the next point, evaluate the objective value and update the posterior model, until we reach the budget of evaluations.

While BO generally performs well in the low-dimensional case, it is well established that it converges more slowly when the dimension is higher. In our problem, the objective function is highly expensive to evaluate due to the complexity of the simulator, so the total budget evaluation tends to be small, say several hundred. To find an eligible point within such a limited budget, we apply a state-of-the-art variant of BO called TuRBO, which adopts a local strategy by maintaining a trust region. We refer to~\cite{eriksson2019scalable} for more details. An interesting observation from our numerical experiments is that TuRBO achieves an efficiency improvement within only several hundred evaluations, although it is originally designed for larger budgets such as thousands of or even tens of thousands of evaluations.

Finally, We discuss the pros and cons of using GP as the surrogate model. GP is not only simple, but also possesses elegant probabilistic properties, and thus we can make statistical inference with the posterior model. Nevertheless, the workload of computing the covariance matrix grows cubically with the number of evaluations, and hence it cannot be scaled to large budget. Moreover, it is often questioned whether this simple model is capable of approximating the complicated objective function, especially if the objective function is rugged. Based on these limitations, surrogate models using various machine learning tools have been developed~\cite{snoek2015scalable,lamperti2018agent}. However, we still decide to use GP as it fits our problem setting. On the one hand, the simulation itself takes so much time that we can only afford a limited total budget, with which computing the covariance matrix is not yet a concern. On the other hand, without the observation noise, the K-S statistic is relatively smooth in the input configuration $\theta$, so GP is capable of approximating this particular objective function. Therefore, the weaknesses of GP are minor here while we could leverage its strengths.

\section{Experimental Results} \label{sec:experiment}

We apply our calibration framework to the Agent-Based Interactive Discrete-Event Simulation (ABIDES) environment~\cite{byrd2020abides} which simulates limit order book (LOB) exchange markets. In this paper, we illustrate our proposed calibration framework with synthetic data. We designate a synthetic ground truth simulator configuration and use it to generate ``real data'' with respect to which we conduct calibration. In such an experimental setting, there is no error due to model assumptions, which makes it easier to assess the performance of our calibration framework.
We further demonstrate our method performance on calibration examples for two-, six- and eighteen-parameter cases. 
In the below experiments, we apply the KS-statistic to the concatenated minute-level mid price return and volume time series samples that result from the market agent interaction directly without prior feature extraction. Please refer to Appendix~\ref{app:experiment} for a more detailed experiment setup, and Appendix~\ref{app:feature_extractor} for a discussion about feature extractors.

\subsection{Limit Order Book Exchange Market Simulator}
The ABIDES model simulates limit order book (LOB) exchange markets with various types of background agents, a NASDAQ-like exchange agent and a simulation kernel managing the flow of time as well as all the agent interactions. More specifically, we consider three types of basic trading agents here. Value agents are designed to simulate the fundamental traders who trade based on their judgment of the fundamental value of the asset. In the simulation system, the fundamental price is modeled as a discrete-time mean-reverting Ornstein-Uhlenbeck process~\cite{byrd2019explaining}. Value agents arrive to the market following a Poisson process and trade a stock depending on whether it is cheap or expensive relative to their own noisy observation of the fundamental price. Besides, noise agents emulate the retail traders who trade on demand without other considerations. Each noise agent arrives to the market at a time uniformly distributed throughout the trading day and place an order of a random size in a random direction. Finally, market maker agents arrive with a constant rate and place limit orders on both sides of the LOB to provide liquidity. The parameters that we aim to calibrate in our experiments include the number, the arrival rate and the minimum/maximum order size of each type of agents as well as the mean value, the mean reversion rate and the volatility of the fundamental.

While the simulator is able to output all the agent activities including bid prices and ask prices, such labeled data are typically unavailable. Thus, we extract minute mid price returns and minute traded volumes from the simulated limit order book, and then concatenate them as the output time series. We only keep the data from 10:00am to 4:00pm. Our goal is to calibrate the parameters of this simulation system only from the output time series sample set.

\subsection{Effectiveness of Plain K-S Statistic with No Feature Extraction}
\label{sec:plain_KS}
\begin{figure}
    \centering
    \begin{subfigure}[b]{0.23\textwidth}
    \centering
    \includegraphics[width=\textwidth]{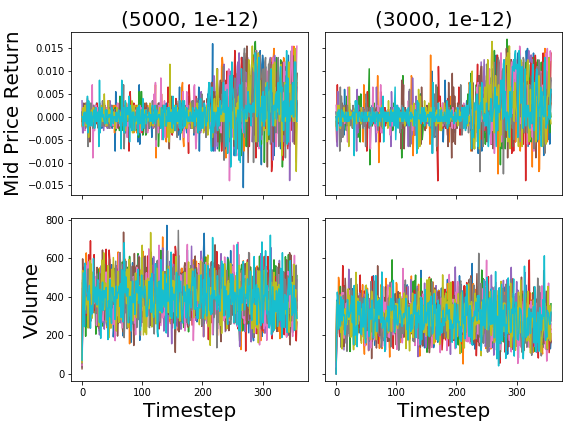}
    \caption{Example 1: Time Series Samples}
    \label{fig:example1_time_series}
    \end{subfigure}
    \begin{subfigure}[b]{0.23\textwidth}
    \centering
    \includegraphics[width=\textwidth]{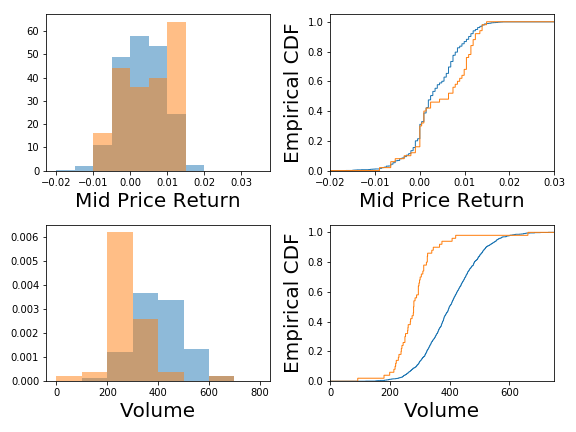}
    \caption{Example 1: Discrepancy in Distributions}
    \label{fig:example1_distribution}
    \end{subfigure}
    \\
    \begin{subfigure}[b]{0.23\textwidth}
    \centering
    \includegraphics[width=\textwidth]{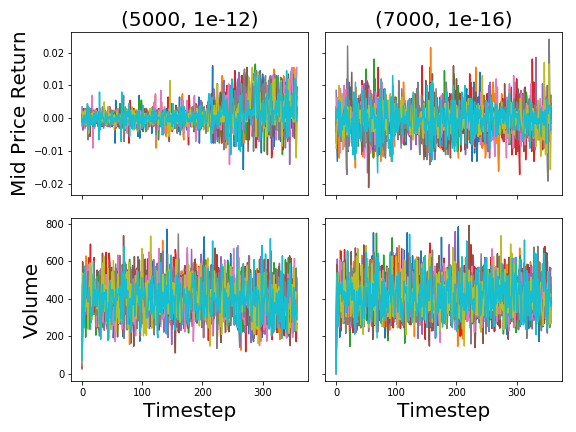}
    \caption{Example 2: Time Series Samples}
    \label{fig:example2_time_series}
    \end{subfigure}
    \begin{subfigure}[b]{0.23\textwidth}
    \centering
    \includegraphics[width=\textwidth]{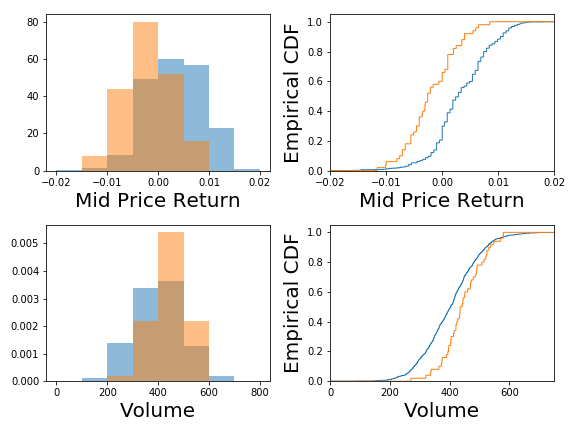}
    \caption{Example 2: Discrepancy in Distributions}
    \label{fig:example2_distribution}
    \end{subfigure}
    \caption{Visualization of the K-S statistic with two examples. (a) and (c) plot 10 time series samples under each configuration. (b) and (d) plot the discrepancy in the distributions.}
\end{figure}

We will verify the effectiveness of using the plain K-S statistic without feature extraction with the two-parameter calibration example, whose setups can be found in Appendix \ref{app:experiment}. First, we use two simple examples to visualize the K-S statistic and show the effectiveness. In the first example, we compare the true configuration $(5000,1e-16)$ with a fake configuration $(3000,1e-16)$. Figure~\ref{fig:example1_time_series} plots 10 time series samples for each configuration. We observe that the mid price return time series distributions look similar while the volume time series seem to come from different distributions. In fact, the fake configuration cannot be rejected by the K-S test applied only on the mid price return time series, but can be rejected only with the volume time series. In Figure~\ref{fig:example1_distribution}, we find the dimensions in the mid price return as well as the volume time series that maximize the one-dimensional K-S statistic and plot the discrepancy in the distributions over these dimensions. It is verified that the discrepancy in the volume distribution is more significant. In the second example, we select another fake configuration $(7000,1e-16)$. In this case, the volume time series look similar while the mid price return time series look obviously different (see Figure~\ref{fig:example2_time_series}). The K-S statistic also coincides with our observation in this example (see Figure~\ref{fig:example2_distribution}). Combining the two examples, we conclude that both price and volume time series are useful in calibration, and the K-S statistic is consistent with intuitive judgments.

\begin{figure}
\centering
\includegraphics[width=0.23\textwidth]{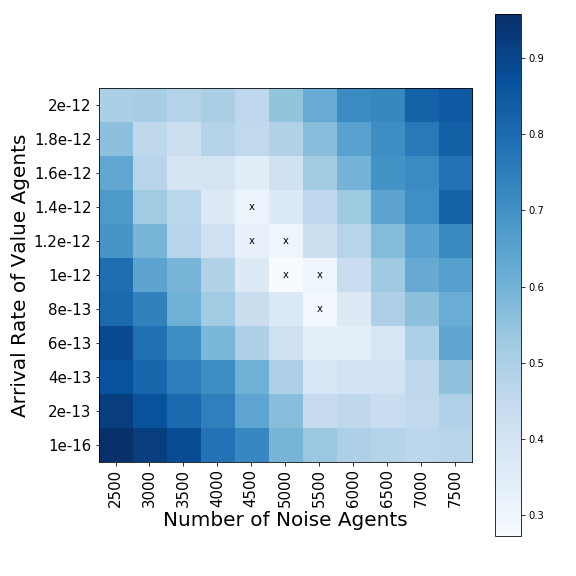}
\caption{Heatmap of K-S statistic without feature extraction with respect to the number of noise agents and the arrival rate of value agents. The grid points with ``x'' marks are in the eligibility set.}
\label{fig:grid_KS_min}
\end{figure}

More systematically, we compute the K-S statistic at some grid points in the parameter space and plot the heatmap in Figure~\ref{fig:grid_KS_min}. Note that the true configuration (5000, 1e-12) achieves the minimum K-S statistic among all the grid points. We also mark the points in the eligibility set. We find that this approximate eligibility set is relatively small compared to the entire space and it looks like an ellipsoid around the truth. Generally, the K-S statistic looks smooth with respect to the parameter value, though there is indeed some randomness. Overall, we decide to directly use the K-S statistic without feature extraction in the experiments.

\subsection{Calibration Performance of BO and Fitting Performance of GP}
\begin{figure}
\centering
\begin{subfigure}[b]{0.23\textwidth}
\centering
\includegraphics[width=\textwidth]{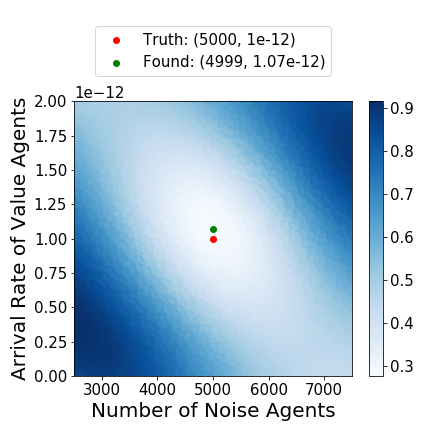}
\caption{Performance of Calibration}
\label{fig:BO_2dim_prediction}
\end{subfigure}
\begin{subfigure}[b]{0.23\textwidth}
\centering
\includegraphics[width=\textwidth]{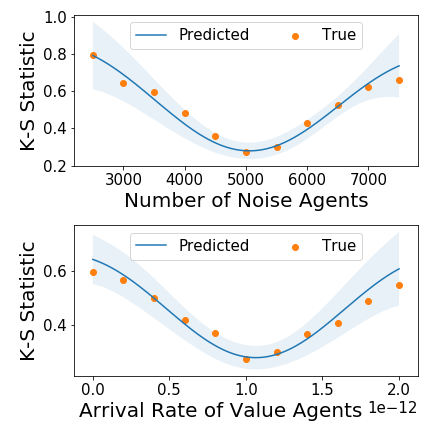}
\caption{Performance of GP Fitting}
\label{fig:BO_2dim_fitting}
\end{subfigure}
\caption{Performance on two-parameter ABIDES calibration example.}
\label{fig:BO_2dim}
\end{figure}

Figure~\ref{fig:BO_2dim} shows the experiment result of applying the BO algorithm on the two-parameter calibration problem. From Figure~\ref{fig:BO_2dim_prediction}, we see that the found optimal point is close to the truth. In particular, it is in the eligibility set, so the calibration is successful. The color represents the predicted K-S statistic by the posterior GP model, which looks similar to the grid search plot (Figure~\ref{fig:grid_KS_min}). To further justify that the posterior GP model is a good fitting for the K-S statistic, we fix one parameter as the truth and perturb the other one, and then compare the true statistic values with the confidence band predicted by the posterior GP. Figure~\ref{fig:BO_2dim_fitting} shows that the band is not wide, especially near the true parameter value. Moreover, the true statistic values lie in the confidence band and basically around the predicted curve. Thus, in this example, GP is able to fit the K-S statistic well.

\subsection{Non-Identifiability Issue}
\begin{figure}
    \centering
    \includegraphics[width=0.3\textwidth]{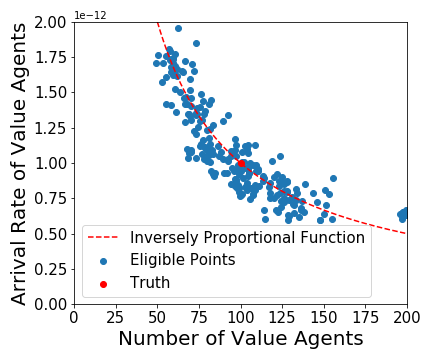}
\caption{Non-identifiability in six-parameter ABIDES calibration example.}
\label{fig:6dim_eligibility_set_subplot}
\end{figure}

In the two-parameter example, the eligibility set is approximately an ellipsoid around the truth, so the problem seems to be identifiable. However, with more parameters incorporated, the non-identifiability issue appears. We collect all the eligible points found in the six-parameter experiments, which forms an approximation to the eligibility set. Figure~\ref{fig:6dim_eligibility_set_subplot} plots the arrival rate against the number of value agents, and we observe that these two parameters are approximately inversely proportional to each other in these eligible points, which makes sense since the overall arrival rate of value agents is the product of these two parameters. 

\subsection{Efficiency Improvement with TuRBO}

\begin{figure}
\centering
\begin{subfigure}[b]{0.23\textwidth}
\centering
\includegraphics[width=\textwidth]{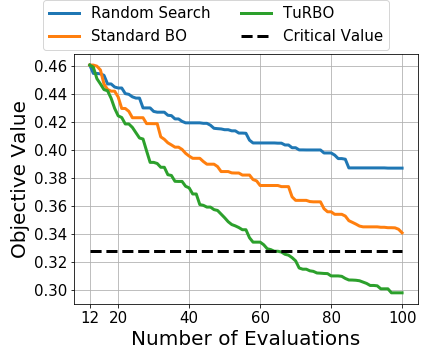}
\caption{Convergence Rate}
\label{fig:BO_6dim_evolution}
\end{subfigure}
\begin{subfigure}[b]{0.23\textwidth}
\centering
\includegraphics[width=\textwidth]{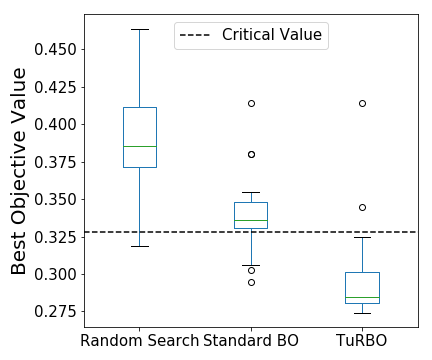}
\caption{Final Optimal Value}
\label{fig:BO_6dim_bestvalue}
\end{subfigure}\\
\begin{subfigure}[b]{0.4\textwidth}
\centering
\begin{tabular}{lr}
    \toprule
    \multicolumn{1}{c}{Method} & \multicolumn{1}{c}{Success Rate} \\
    \midrule
    Random Search & 5\% \\
    Standard BO & 20\% \\
    TuRBO & 90\% \\
    \bottomrule
    \end{tabular}
    \caption{Success Rate}
    \label{fig:BO_6dim_successrate}
\end{subfigure}
\caption{Comparison of random search, standard BO and TuRBO on six-parameter ABIDES calibration example.}
\label{fig:BO_6dim}
\end{figure}

\begin{figure}
\centering
\begin{subfigure}[b]{0.23\textwidth}
\centering
\includegraphics[width=\textwidth]{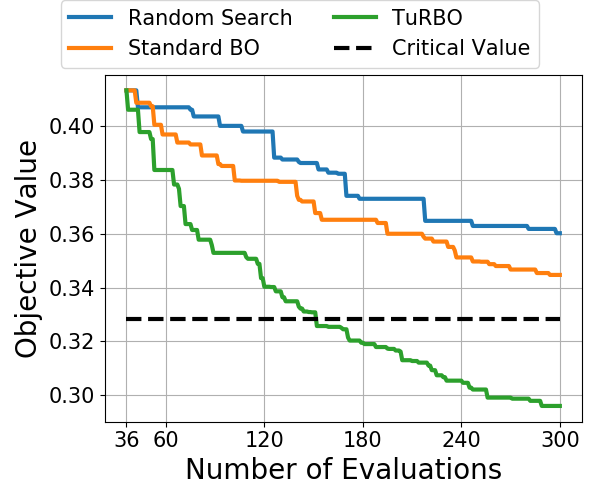}
\caption{Convergence Rate}
\label{fig:BO_18dim_evolution}
\end{subfigure}
\begin{subfigure}[b]{0.23\textwidth}
\centering
\includegraphics[width=\textwidth]{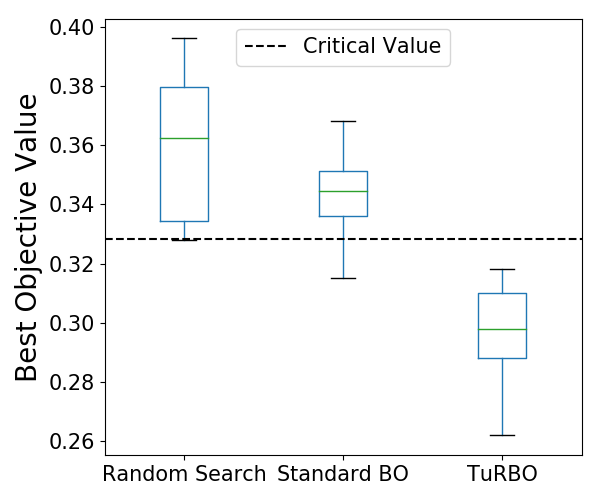}
\caption{Final Optimal Value}
\label{fig:BO_18dim_bestvalue}
\end{subfigure}\\
\begin{subfigure}[b]{0.4\textwidth}
\centering
\begin{tabular}{lr}
    \toprule
    \multicolumn{1}{c}{Method} & \multicolumn{1}{c}{Success Rate} \\
    \midrule
    Random Search & 10\% \\
    Standard BO & 10\% \\
    TuRBO & 100\% \\
    \bottomrule

    \end{tabular}
    \caption{Success Rate}
    \label{fig:BO_18dim_successrate}
\end{subfigure}
\caption{Comparison of random search, standard BO and TuRBO on eighteen-parameter ABIDES calibration example.}
\label{fig:BO_18dim}
\end{figure}

Although the performance in the two-parameter calibration example is satisfactory, we observe that the standard BO algorithm converges more slowly as the number of parameters increases. In the six-parameter or eighteen-parameter experiments, we compare the efficiency of random search, standard BO and TuRBO with 20 or 10 random seeds to reduce randomness. Figures~\ref{fig:BO_6dim} and~\ref{fig:BO_18dim} show the experiment results. Figure~\ref{fig:BO_6dim_evolution} shows how the best objective value evolves as the number of evaluations increases, where we take an average over 20 experiments. Figure~\ref{fig:BO_6dim_bestvalue} compares the finally found best objective values under the 20 random seeds for each method. Moreover, if we regard finding an eligible point within 100 evaluations as a success, then Table~\ref{fig:BO_6dim_successrate} shows the success rate of each method. Overall, in terms of both convergence rate and success rate, TuRBO is the best while random search is the worst. Similarly, Figure~\ref{fig:BO_18dim} further supports that TuRBO still greatly outperforms the other two methods in such a high-dimensional problem.

\section{Conclusion} \label{sec:conclusion}

In summary, we propose an efficient framework to calibrate multi-agent simulators from the output series. Specifically, we use the novel concept of eligibility set to bypass the possible non-identifiability issue in large-scale simulation systems, use the K-S statistic generalized to high-dimensional cases as the distance metric and optimize the distance metric with BO or TuRBO. Our numerical experiments consistently show that our framework with TuRBO is able to recover an eligible configuration with highly limited budget of evaluations, which achieves a substantial improvement in efficiency compared to the random search method.

In the future, we plan to extend this work towards several directions. First, we can apply the calibration framework to real market data and test the effectiveness of our framework as well as the reliability of our simulator. Second, we can try the calibration framework on higher-dimensional calibration problems to investigate its limit. Third, we can reduce the conservativeness of the K-S statistic by developing useful feature extractors. 

\paragraph{Disclaimer}This paper was prepared for informational purposes [“in part” if the work is collaborative with external partners] by the Artificial Intelligence Research group of JPMorgan Chase and Co and its affiliates (“J.P. Morgan”), and is not a product of the Research Department of J.P. Morgan.  J.P. Morgan makes no representation and warranty whatsoever and disclaims all liability, for the completeness, accuracy or reliability of the information contained herein.  This document is not intended as investment research or investment advice, or a recommendation, offer or solicitation for the purchase or sale of any security, financial instrument, financial product or service, or to be used in any way for evaluating the merits of participating in any transaction, and shall not constitute a solicitation under any jurisdiction or to any person, if such solicitation under such jurisdiction or to such person would be unlawful.

\begin{acks}
This work was done while Yuanlu Bai was working as a summer intern with J.P. Morgan AI Research, under supervision of her academic advisor Henry Lam and internship managers Svitlana Vyetrenko and Tucker Balch. We also acknowledge support from the J. P. Morgan Chase Faculty Research Award.
\end{acks}

\bibliographystyle{ACM-Reference-Format}
\bibliography{bibliography}

\appendix

\section{Experiment Setup}
\label{app:experiment}

We pick a ``true'' configuration and generate 1000 real time series samples using this configuration and different random seeds. Then for each candidate configuration in the parameter space, we generate 50 time series samples to compute the K-S statistic. The significance level $\alpha$ is chosen as 0.05. That is, the eligibility set is an approximate $95\%$ confidence region. 

To assess the calibration performance under different dimensions, we gradually increase the number of flexible parameters while fixing the other parameters to be same as the true configuration. In the experiments, we respectively consider calibrating two, six and eighteen parameters. The full parameter list used in the eighteen-parameter example is in Table~\ref{tab:param_18dim}. In the two- and six-parameter examples, we respectively calibrate the subset (num\_noise, lambda\_a\_1) and (num\_value\_1, num\_noise, lambda\_a\_1, r\_bar, kappa, fund\_vol).

In the two-parameter example, to run the BO algorithm, we set the initial budget $n_{init}$ as 10 and the total budget $n_{total}$ as 100. We use the commonly-used RBF kernel as the GP kernel. White noise is added to the kernel considering that our distance metric is actually noisy. We use a mixture of expected improvement (EI), probability of improvement (PI) and lower confidence bound (LCB) as the acquisition function, which is the default choice of the skopt package.

In the six- and eighteen-parameter calibration example, we compare the performance of three algorithms: random search, standard BO and TuRBO. The total budget is fixed as 100 and 300 respectively. Random search is simply randomly sampling points in the parameter space and evaluating the objective values. We try to keep the settings in standard BO and TuRBO identical as much as possible to be fair. For both algorithms, we set the initial budget as $2d$ where $d$ is the number of parameters, use the Mat\'{e}rn-5/2 kernel with white noise as the GP covariance kernel and choose Thompson sampling (TS) with respectively 500 and 3600 candidate points as the acquisition function. Specifically for TuRBO, the hyperparameters are chosen as: number of trust regions $m=1$, batch size $q=1$, $\tau_{succ}=1$, $\tau_{fail}=d$, $L_{min}=2^{-7}$, $L_{max}=1.6$, $L_{init}=0.8$. We refer to~\cite{eriksson2019scalable} for more details about the hyperparameters and we note that our choice is mostly the same as their recommendation, except a little adjustment due to our much more limited budget of evaluations than theirs. To reduce randomness of experiments, we test the three algorithms with respectively 20 and 10 different random seeds. Under the same seed, the sampled initial points and the corresponding K-S statistic values are the same for each algorithm, so they ``start off'' with the same information about the objective function.

\begin{tiny}
\begin{table}
  \centering
  \caption{Parameters to calibrate in eighteen-parameter calibration example.}
    \begin{tabular}{cccc}
    \toprule
    Parameter Symbol & Meaning & True Value & Range  \\
    \midrule
    num\_value\_1 & Number of Type 1 Value Agents & 100   & [0,200] \\
    lambda\_a\_1 & Arrival Rate of Type 1 Value Agents & 1e-12 & [1e-16,2e-12] \\
    min\_size\_value\_1 & Minimum Order Size of Type 1 Value Agents & 20    & [6,34] \\
    max\_size\_value\_1 & \multicolumn{1}{l}{Maximum Order Size of Type 1 Value Agents} & 50    & [36,64] \\
    num\_value\_2 & Number of Type 2 Value Agents & 0     & [0,50] \\
    lambda\_a\_2 & Arrival Rate of Type 2 Value Agents & 1e-12 & [1e-16,2e-12] \\
    min\_size\_value\_2 & Minimum Order Size of Type 2 Value Agents & 100   & [76,124] \\
    max\_size\_value\_2 & \multicolumn{1}{l}{Maximum Order Size of Type 2 Value Agents} & 150   & [126,174] \\
    num\_value\_3 & Number of Type 3 Value Agents & 0     & [0,50] \\
    lambda\_a\_3 & Arrival Rate of Type 3 Value Agents & 1e-12 & [1e-16,2e-12] \\
    min\_size\_value\_3 & Minimum Order Size of Type 3 Value Agents & 200   & [176,224] \\
    max\_size\_value\_3 & \multicolumn{1}{l}{Maximum Order Size of Type 3 Value Agents} & 250   & [226,274] \\
    num\_noise & Number of Noise Agents & 5000  & [2500,7500] \\
    min\_size\_noise & Minimum Order Size of Noise Agents & 20    & [6,34] \\
    max\_size\_noise & Maximum Order Size of Noise Agents & 50    & [36,64] \\
    r\_bar & Mean of Fundamental & 1e5   & [1e2,2e5] \\
    kappa & Mean Reversion Rate of Fundamental & 1.67e-12 & [1e-16,3e-12] \\
    fund\_vol & Fundamental Volatility & 1e-4  & [1e-8,1] \\
    \bottomrule
    \end{tabular}%
  \label{tab:param_18dim}%
\end{table}%
\end{tiny}

\section{Discussion about Feature Extractors}
\label{app:feature_extractor}
With the two-parameter ABIDES calibration example, we investigate how to choose the feature extractor $f$. In the experiments, we adopt the simplest idea which is not to apply any feature extraction at all. We will discuss more possible selections of feature extractors.

\subsection{Conservativeness of K-S statistic with High-Dimensional Time Series}
\label{app:conservativeness}
\begin{figure}
\centering
\includegraphics[width=0.23\textwidth]{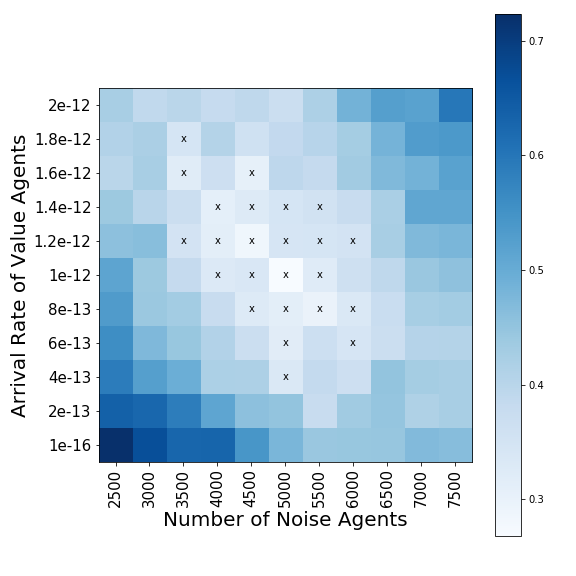}
\caption{Heatmap of K-S statistic without feature extraction using 10 second data. The grid points with ``x'' marks are in the eligibility set.}
\label{fig:grid_KS_10s}
\end{figure}

We have shown the effectiveness of the K-S statistic used in our experiments. However, we note that the K-S statistic can be conservative especially when the length of time series is too large due to the nature of Bonferroni correction. Compared with the minute data, we try to increase the sampling frequency to 10 seconds and hence the length of the time series increases to over 4,000. We plot the heatmap in Figure~\ref{fig:grid_KS_10s} and we observe that the eligibility set is much larger. Some configurations could have been rejected with minute data but fail to be rejected with the 10 seconds data. In this case, applying a suitable feature extractor before computing the K-S statistic can be helpful. 

\subsection{Feature Extractors using Learning-Based Methods}
\label{app:learning_based}
Various learning-based feature extractors have been developed. We skip the detailed discussion here and refer to~\cite{bai2021calibrating,masgan} for feature extractors of high-dimensional financial time series using autoencoder, GAN or WGAN, which can effectively reduce the dimension and hence the conservativeness of the K-S statistic. However, while these machine learning models are flexible and powerful, they usually lack interpretability and also they need to be retrained whenever they are applied to a different problem setting.

\subsection{Feature Extractors using Stylized Fact Metrics}
\label{app:stylized_fact}

\begin{figure}[htbp]
\centering
\begin{subfigure}[b]{0.23\textwidth}
\centering
\includegraphics[width=\textwidth]{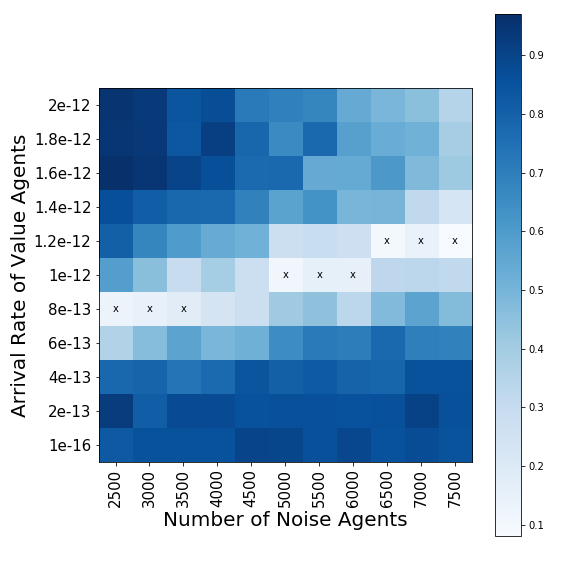}
\caption{Autocorrelation}
\label{fig:grid_KS_autocorrelation}
\end{subfigure}
\begin{subfigure}[b]{0.23\textwidth}
\centering
\includegraphics[width=\textwidth]{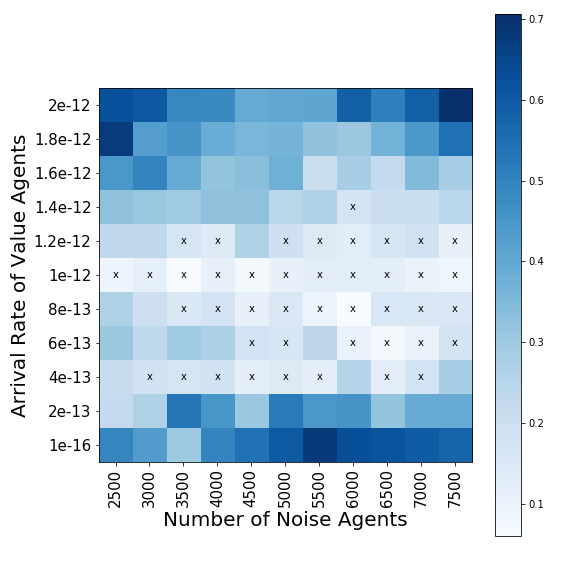}
\caption{Kurtosis}
\label{fig:grid_KS_kurtosis}
\end{subfigure}\\
\begin{subfigure}[b]{0.23\textwidth}
\centering
\includegraphics[width=\textwidth]{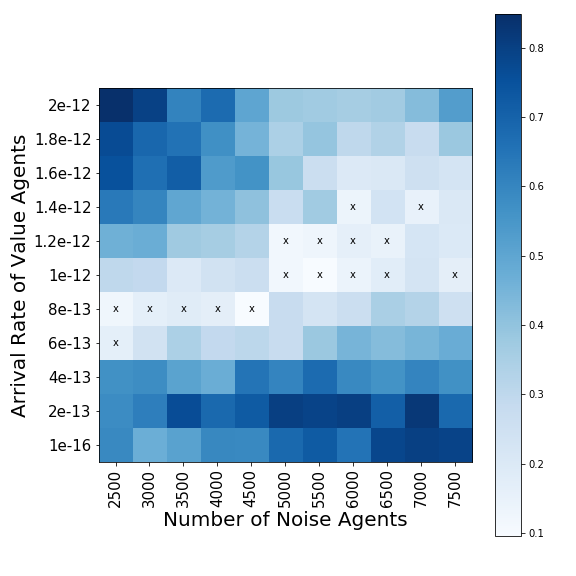}
\caption{Volatility Clustering}
\label{fig:grid_KS_volatility_clustering}
\end{subfigure}
\begin{subfigure}[b]{0.23\textwidth}
\centering
\includegraphics[width=\textwidth]{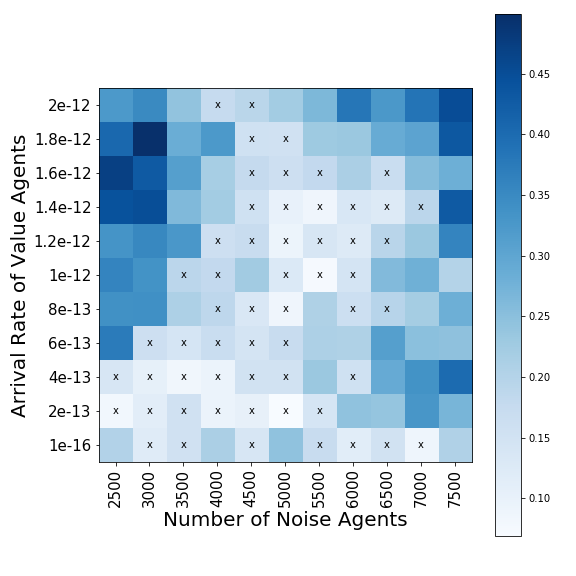}
\caption{Volume Volatility Correlation}
\label{fig:grid_KS_volume_volatility_correlation}
\end{subfigure}\\
\begin{subfigure}[b]{0.23\textwidth}
\centering
\includegraphics[width=\textwidth]{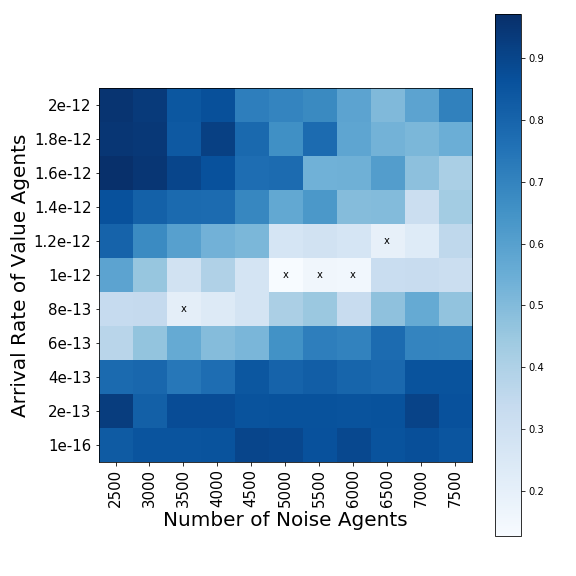}
\caption{Combining Four Metrics}
\label{fig:grid_KS_all_metrics}
\end{subfigure}
\caption{Heatmap of K-S statistic using stylized fact metrics as the feature extractor.}
\label{fig:grid_KS_realism_metrics}
\end{figure}

Finally, we investigate whether we can use stylized fact metrics with physical meanings to construct the feature extractor. We consider four metrics discussed in~\cite{vyetrenko2019get}: autocorrelation, kurtosis, volatility clustering and volume volatility correlation. To be specific, we denote the minute log return as $r_t$ and the volume as $V_t$. Then they are respectively the correlation between $r_t$ and $r_{t+1}$, the kurtosis of $r_t$, the correlation between $r_t^2$ and $r_{t+1}^2$, and the correlation coefficient of $V_t$ against $|r_t|$. With each time series sample, we compute
an estimate for each metric, and then use these estimates as the extracted features. Figure~\ref{fig:grid_KS_realism_metrics} shows the heatmaps of the K-S statistic
using each metric alone and also the corresponding eligible grid points. We see that for each of them, the K-S statistic seems not to vary with the parameter value regularly. If we combine the four metrics as a single feature extractor (i.e., $f:\R^O\to\R^4$), then as
shown in Figure~\ref{fig:grid_KS_all_metrics}, the eligibility set looks small, but there seem to be a lot of local minima, which will increase the difficulty of optimization. Based on our investigation, despite their usefulness in general, the stylized fact metrics may not be sufficient to work as the feature extractor since too much information is lost when
the entire time series is summarized into a few numbers.

\newpage


\end{document}